\documentclass[journal,twoside,web]{ieeecolor}
\usepackage{tmi}
\usepackage{cite}
\usepackage{amsmath,amssymb,amsfonts}
\usepackage{algpseudocode}
\usepackage{graphicx}
\usepackage{textcomp}
\usepackage{array, multirow, hhline, lscape, chngcntr, pdfpages, float}
\usepackage{booktabs}
\usepackage{url}
\usepackage{hyperref}

\usepackage{colortbl}
\definecolor{mygray}{gray}{0.9}
\usepackage{algorithm}

\usepackage{tikz}

\newcommand{\x}{\mathbf{X}}
\newcommand{\y}{\mathbf{Y}}

\newcommand{\ie}{{\em i.e.}}
\newcommand{\eg}{{\em e.g.}}
\newcommand{\etal}{{\em et al.}}

\def\BibTeX{{\rm B\kern-.05em{\sc i\kern-.025em b}\kern-.08em
    T\kern-.1667em\lower.7ex\hbox{E}\kern-.125emX}}
\begin{document}
\title{DyMix: Dynamic Frequency Mixup Scheduler based Unsupervised Domain Adaptation for Enhancing Alzheimer's Disease Identification}
\author{Yooseung Shin, Kwanseok Oh, and Heung-Il Suk \IEEEmembership{Senior Member, IEEE}
\thanks{This work was supported by the National Research Foundation of Korea (NRF) grant funded by the Korea government (No. 2022R1A4A1033856) and the Institute of Information \& communications Technology Planning \& Evaluation (IITP) grant funded by the Korea government(MSIT) No. RS-2022-II220959 ((Part 2) Few-Shot Learning of Causal Inference in Vision and Language for Decision Making) and (No. RS-2019-II190079, Artificial Intelligence Graduate School Program(Korea University)).}
\thanks{Y. Shin and K. Oh are with the Department of Artificial Intelligence, Korea University, Seoul 02841, Republic of Korea (e-mail: {usxxng, ksohh}@korea.ac.kr)}
\thanks{H.-I. Suk is with the Department of Artificial Intelligence, Korea University, Seoul 02841, Republic of Korea and also with the Department of Brain and Cognitive Engineering, Korea University, Seoul 02841, Republic of Korea (e-mail: hisuk@korea.ac.kr, Corresponding author).}
\thanks{Y. Shin and K. Oh have contributed equally to this work.}}

\maketitle

\begin{abstract}
Advances in deep learning (DL)-based models for brain image analysis have significantly enhanced the accuracy of Alzheimer's disease (AD) diagnosis, allowing for more timely interventions. Despite these advancements, most current DL models suffer from performance degradation when inferring on unseen domain data owing to the variations in data distributions, a phenomenon known as domain shift. To address this challenge, we propose a novel approach called the dynamic frequency mixup scheduler (DyMix) for unsupervised domain adaptation. Contrary to the conventional mixup technique, which involves simple linear interpolations between predefined data points from the frequency space, our proposed DyMix dynamically adjusts the magnitude of the frequency regions being mixed from the source and target domains. Such an adaptive strategy optimizes the model’s capacity to deal with domain variability, thereby enhancing its generalizability across the target domain. In addition, we incorporate additional strategies to further enforce the model's robustness against domain shifts, including leveraging amplitude-phase recombination to ensure resilience to intensity variations and applying self-adversarial learning to derive domain-invariant feature representations. Experimental results on two benchmark datasets quantitatively and qualitatively validated the effectiveness of our DyMix in that we demonstrated its outstanding performance in AD diagnosis compared to state-of-the-art methods. The code is available at: \url{https://github.com/ku-milab/DyMix}.
\end{abstract}


\section{Introduction}
\label{sec:introduction}
\IEEEPARstart{P}{recise} identification of prevalent brain disorders is essential for timely intervention and treatment. It also plays a significant role in advancing neuroscience research on therapeutic development. Of diverse brain imaging tools, structural magnetic resonance imaging (sMRI) is a pivotal tool for providing detailed images of brain anatomy \cite{frisoni2010clinical}, enabling researchers and clinicians to detect abnormalities associated with health conditions, such as Alzheimer's disease (AD) or its prodromal stage, known as mild cognitive impairment (MCI) \cite{brookmeyer2007forecasting}. AD, an irreversible neurodegenerative disease, progressively leads to cognitive decline and severe memory impairment and there is currently no apparent cure for AD \cite{alzheimer20192019} is known. Therefore, early and accurate identification is critical for delaying the progression of the disease and improving patient care.

\begin{figure}[!t]
    \centering
    \includegraphics[width=1.0\columnwidth]{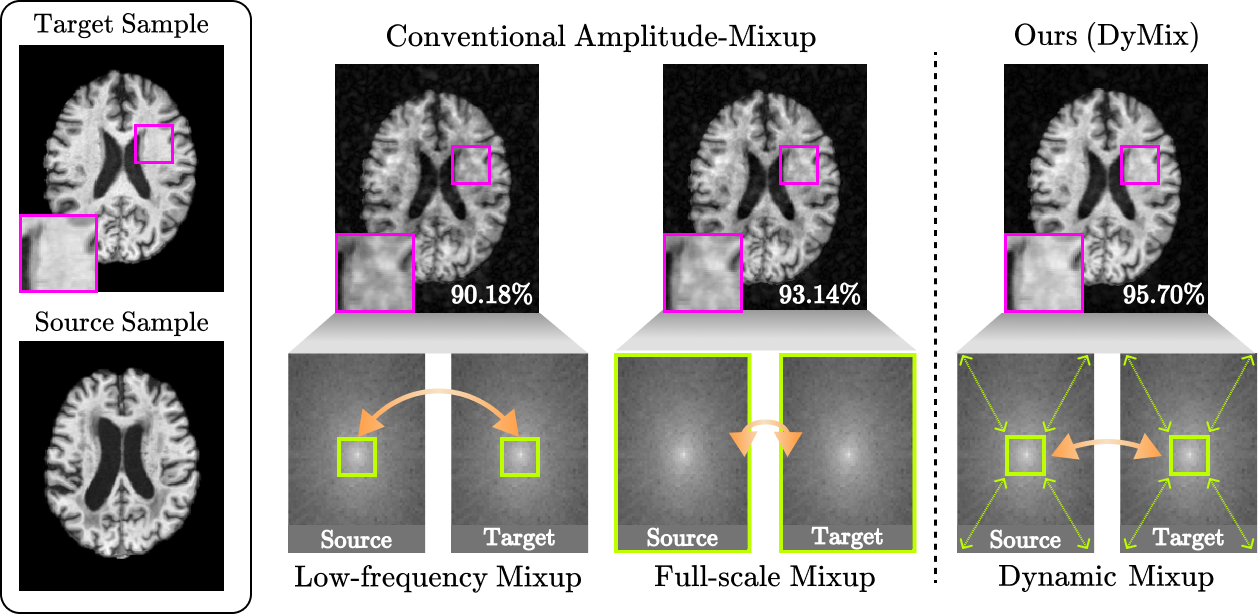}
    \caption{The primary difference between conventional amplitude mixup techniques and our proposed DyMix. Here, the posterior probabilities in each manipulated image denote the classification accuracy derived from the trained model using their respective augmentation strategies.}
    \label{fig:intro_fig}
\end{figure}

Based on sMRI data curated from diverse sites/institutions, various learning-based approaches have devoted their efforts to enhancing AD diagnostic accuracy and reliability \cite{zhao2023conventional,khan2021machine}. Among these, advances in deep learning (DL)-based methods have revolutionized the field \cite{zhang2020survey} by automatically extracting and learning intricate features for profound atrophies caused by AD.  
However, such success of DL methods is heavily contingent upon an underlying premise that the training data (\ie, source domain) and test data (\ie, target domain) phase have arranged to a uniform data distribution, that is, an independent and identically distributed assumption. If such an indispensable assumption is slightly unsatisfied or even violated, the DL model's diagnostic performance may deteriorate severely—a phenomenon known as the domain shift \cite{ben2006analysis}. In medical imaging, most domain shifts can arise from differences in data acquisition institutions, variations in scanner protocols, or other medical factors, all of which can lead to domain discrepancies between the source and target data domains.

Unsupervised domain adaptation (UDA) has been introduced to align the distributions of source and target data domains to alleviate the impact of domain shifts across different datasets. The strategy of UDA methods typically transfers knowledge from the labeled source data to the target data without using target labels \cite{wilson2020survey}. In this context, domain-adversarial training of neural networks (DANN) \cite{ganin2016domain}, a broadly used method in medical imaging, leverages UDA-based adversarial learning to minimize domain discrepancies. Deep correlation alignment (Deep-CORAL) \cite{sun2017correlation} focuses on aligning the second-order statistical properties between source and target distributions, effectively reducing the need for target labels. Additionally, advanced manners such as an attention-guided deep domain adaptation (AD$^{2}$A) \cite{guan2021multi} and a deep prototype-guided multi-scale domain adaptation (PMDA) \cite{cai2023prototype} introduce more specialized mechanisms, including attention-guided strategies and prototype-guided multiscale adaptation, to further refine feature alignment and tackle issues such as data imbalance. While these UDA methods have made significant strides in addressing domain shifts, particularly by targeting and transforming local regions within the spatial domain of images, they often fall short of capturing the broader context of spatial patterns.
This limitation can be particularly detrimental in medical imaging, where morphological variations in global structures often play a crucial role in accurate diagnosis. Moreover, conventional UDA methods tend to prioritize target domain adaptation, which can lead to less rigorous pretraining of a model in the source domain. Such a drawback becomes pronounced when the source data are either imbalanced or insufficient, causing the source classifier to inevitably struggle with not only with extracting semantically meaningful representations but also with adapting to new or diverse data in the target domain.

To circumvent these challenges, recent research has explored handling the frequency domain via Fourier transformation \cite{nussbaumer1982fast} as an alternative approach to domain alignment \cite{yang2020fda,hu2022domain,ge2023unsupervised,oh2024fiesta}. 
Through the Fourier transformation, an image is decomposed into its two constituent frequencies—amplitude and phase components—where the amplitude component contains the image textures, such as contrast and brightness, and the phase component represents the image structural patterns, such as the overall appearance and object boundaries. Leveraging these inherent characteristics, Fourier-based UDA methods have improved performance by adopting a straightforward manner that involves manipulating a certain portion of the low-frequency spectrum within the amplitude to conduct texture-related image transformations. However, these approaches are confined by their exclusive focus on manually predefined low-frequency regions, which often results in the neglect of equally essential high-frequency properties that are equally essential. From this perspective, Shin \etal \cite{shin2023frequency} attempted full-scale frequency mixing, in which the entire range of frequencies is exploited for image manipulation. While this approach provides a more comprehensive alignment, it still suffers from identifying the optimal frequency regions for maximizing performance. As the distinction between meaningful domain-specific details and domain-irrelevant noise could be subtle and context-dependent, relying solely on predefined region manipulation of either certain low-frequency or full-frequency regions may not yield the best results, as illustrated in Fig.~\ref{fig:intro_fig}. Consequently, \textit{it is necessary to dynamically identify and adjust the optimal magnitude of frequency regions throughout the training process to ensure that the most relevant frequency information is utilized for effective domain adaptation}.

Building upon these premises, we propose a dynamic frequency mixup scheduler (DyMix), a novel approach designed to automatically identify and blend the optimal regions in the amplitude component for dynamic frequency manipulation. DyMix leverages the mixup technique to combine the amplitudes from both the source and target domains \cite{zhang2017mixup}, aiming to enhance UDA performance. To this end, the proposed method consists of two fundamental steps: (i) pretraining to learn invariant feature representations and (ii) domain adaptation via dynamic frequency manipulation. 
In the pretraining step, we employ the Amplitude-Phase Recombination \cite{chen2021amplitude} to generate intensity-transformed images within the source domain. This involves recombining the amplitude spectrum from the intensity-transformed source image with the phase information from the original source image, thereby effectively generating new representations for increasing data diversity. To further reinforce the model’s robustness, we incorporate self-adversarial learning \cite{zhou2023self} to assist the model in deriving a semantic representation that is invariant to intensity-related changes. As a result, the model is better equipped to handle the variability between the source and target domains, thereby setting a solid foundation for the subsequent dynamic frequency manipulations during the domain-adaptation phase. In the adaptation step, the proposed DyMix is employed to produce a novel amplitude-mixed target image. Here, the pretrained model, which has been exclusively trained on the source domain data, is used to facilitate domain adaptation. Afterward, DyMix dynamically adjusts the amplitude spectrum by gradually increasing or decreasing the boundary magnitude of the amplitude region whenever the evaluation score plateaus during the adaptation phase, ensuring that the optimal frequency regions are selected to improve UDA performance. In this way, our proposed method using DyMix provides a robust and adaptive solution to the challenges posed by domain variability by effectively integrating low-level statistics from the target domain while preserving those from the source domain. Accordingly, the main contributions of this work are as follows:
 
\begin{itemize}
    \item We propose a novel dynamic frequency mixup scheduler (DyMix) that dynamically adjusts the boundary magnitude of manipulation regions within the amplitude component to maximize the UDA performance.
    \item We enhance the generalizability of our approach by incorporating a pretraining step that leverages self-adversarial learning and frequency manipulation to transform the intensity-shifted source domain adaptively, facilitating more robust domain adaptation.
    \item We validate the effectiveness of our DyMix via comprehensive quantitative and qualitative experiments conducted on two benchmark datasets for brain disease classifications: the Alzheimer’s Disease Neuroimaging Initiative (ADNI) \cite{mueller2005alzheimer} and the Australian Imaging Biomarkers and Lifestyle Study of Aging (AIBL) \cite{rowe2010amyloid} datasets.
\end{itemize}

\section{Preliminary: Fourier Transformation}
Before delving into the details of the proposed method, we first discuss the fundamental concepts and formulation needed to understand the Fourier transformation (FT), as it plays a crucial role in developing our approach. Specifically, we revisit how FT extracts amplitude and phase components from an image in the spatial domain. Given a the three-dimensional (3D) input $\x\in \mathbb{R}^{\text{H}\times \text{W}\times \text{D}\times 1}$, the formulation of FT for the 3D input $\x$ can be defined as follows:
\begin{equation}
    \mathcal{F}(\x)=\sum_{h=0}^{H-1} \sum_{w=0}^{W-1} \sum_{d=0}^{D-1} \x(h, w, d)\cdot e^{-j 2 \pi\left(\frac{h}{H} x+\frac{w}{W} y +\frac{d}{D} z\right)}.
\end{equation}
Here, $x$, $y$, and $z$ represent the frequency variables corresponding to the $h$, $w$, and $d$ spatial dimensions, respectively, and $\mathcal{F}(\cdot)$ indicates the fast FT (FFT) \cite{nussbaumer1982fast}.

In this way, the amplitude $\mathcal{A}(\x)$ and phase $\mathcal{P}(\x)$ components are derived from 3D input $\x$ as shown below:
\begin{equation}
    \mathcal{A}(\x)=\sqrt{R^2(\x)(x, y, z)+I^2(\x)(x, y, z)},\label{eq:amp}
\end{equation}
\begin{equation}
    \mathcal{P}(\x)=\arctan \left[\frac{I(\x)(x, y, z)}{R(\x)(x, y, z)}\right],\label{eq:pha}
\end{equation}
where $R(\x)$ and $I(\x)$ denote the real and imaginary parts of the $\mathcal{F}(\x)$, respectively. The inverse FFT (iFFT), denoted by $\mathcal{F}^{-1}(\cdot)$, is used to convert spectral signals, including the amplitude and phase, from the frequency domain reverse in the spatial domain as $\x=\mathcal{F}^{-1}(\mathcal{A}(\x), \mathcal{P}(\x))$. To simplify the remaining sections, the FFT $\mathcal{F}(\cdot)$ and iFFT $\mathcal{F}^{-1}(\cdot)$ are applied with a shift operator that multiplies the amplitude and phase spectra by $(-1)^{x+y+z}$, ensuring that the low-frequency components are centered.

\section{Proposed Method}
\label{sec:method}
{The objective of our proposed method is to train a brain disease classification model using both the source domain $\mathcal{D}_{\text{s}}$ and target domain $\mathcal{D}_{\text{t}}$ so that it can perform effectively in unseen target domains. Specifically, $\{\x_{\text{s}}^i, \y_{\text{s}}^i\}_{i=1}^{N_\text{s}}$ denotes the set of $N_\text{s}$ source data and their counterpart category label in the source domain $\mathcal{D}_{\text{s}}$, while the target domain $\mathcal{D}_{\text{t}}$ consists of the set of $N_{\text{t}}$ unlabeled target data, denoted as $\{\x_{\text{t}}^i\}_{i=1}^{N_{\text{t}}}$. 

To achieve this goal, our framework includes two key steps: (i) pretraining for invariant feature representation learning and (ii) domain adaptation using dynamic frequency manipulation. As illustrated in Fig.~\ref{overall_framework}, the first step involves employing strategies of amplitude-phase recombination \cite{chen2021amplitude} and self-adversarial learning \cite{zhou2023self}, which helps the model become less sensitive to variations among various manipulated image properties. For the second step, the proposed DyMix properly adjusts the frequency regions during training to ensure optimal adaptation. This step involves using a variant amplitude mixup to dynamically blend frequency components from both the source and target domains, producing semantic representations that bridge the gap between the two domains.

\begin{figure*}
    \centerline{\includegraphics[width=\textwidth]{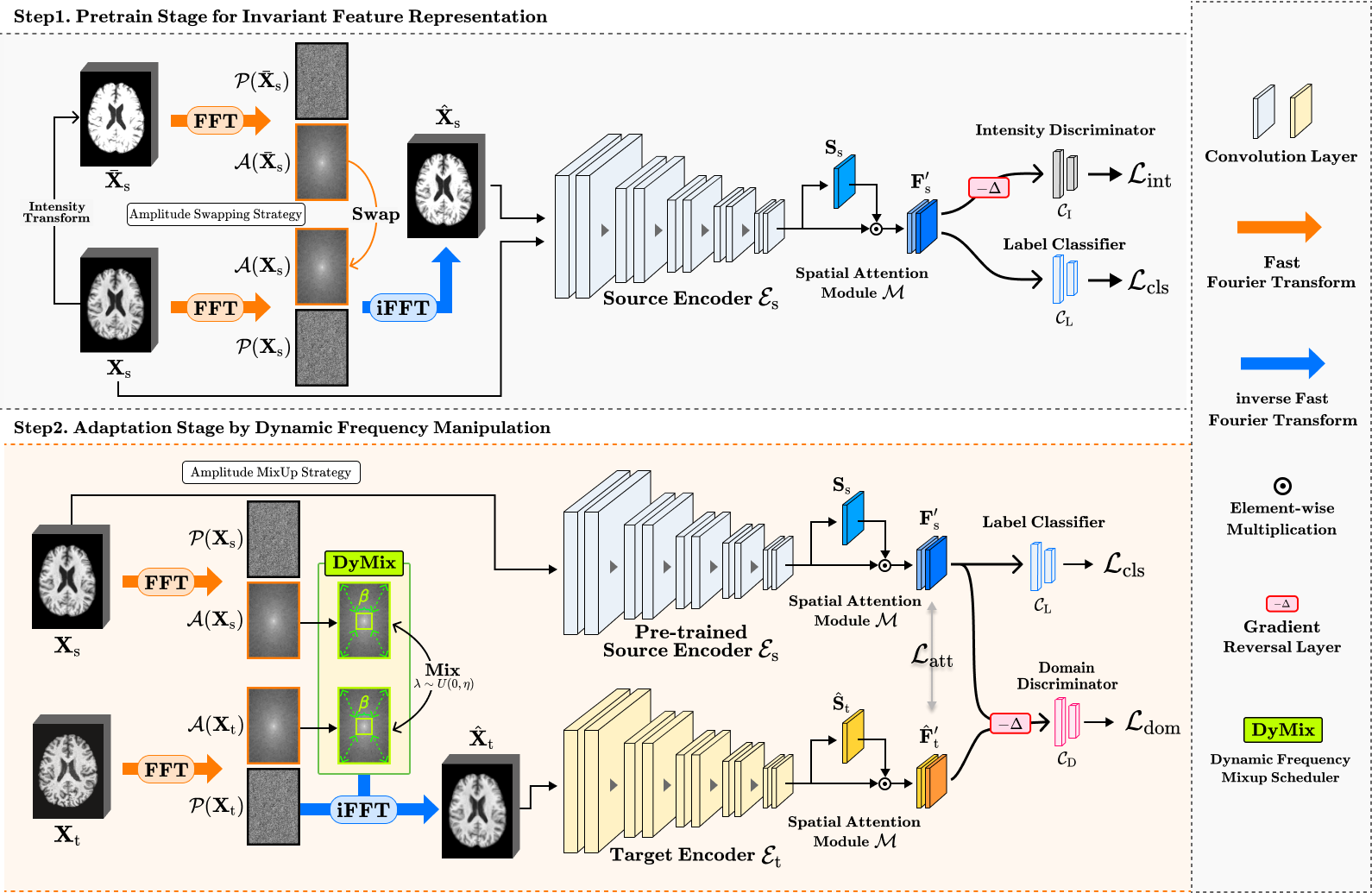}}
    \caption{The overall framework of our proposed method consists of two main steps: (i) the pretraining stage for invariant feature representation and (ii) the adaptation stage by dynamic frequency manipulation. This framework ensures a robust approach to learning and adaptation across different domains.}
    \label{overall_framework}
\end{figure*}

\subsection{Pretraining for Invariant Feature Representation}

\subsubsection{Data Manipulation using Amplitude-Phase Recombination}
Given the source domain dataset $\{\x_{\text{s}}^i, \y_{\text{s}}^i\}_{i=1}^{N_\text{s}}$, we first utilize the RandomBiasField (RBF) \cite{perez2021torchio} transformation along with the source image $\x_{\text{s}}$ to produce an intensity-transformed source image as $\bar{\x}_{\text{s}} = \operatorname{RBF}(\x_{\text{s}})$. Each $\x_{\text{s}}$ and $\bar{\x}_{\text{s}}$ is subsequently decomposed using Eq. \eqref{eq:amp} and Eq. \eqref{eq:pha} to derive the amplitude $\mathcal{A}(\x_{\text{s}}), \mathcal{A}(\bar{\x}_{\text{s}})$ and phase $\mathcal{P}(\x_{\text{s}}), \mathcal{P}(\bar{\x}_{\text{s}})$ components. For conducting the manipulation in frequency space, we employ the Amplitude-Phase Recombination \cite{chen2021amplitude} based on the swapping strategy, and perform the iFFT $\mathcal{F}^{-1}$ to obtain the reconstructed intensity-shifted source image $\hat{\x}_{\text{s}}$ as follows: 
\begin{equation}\label{d_is}
\hat{\x}_{\text{s}} = \mathcal{F}^{-1}(\mathcal{A}(\bar{\x}_{\text{s}}), \mathcal{P}(\x_{\text{s}})).
\end{equation}
Such manipulation produces an intensity-shifted source domain, which retains the semantic characteristics of the original source domain while incorporating different intensity distributions.

\subsubsection{Spatial Attention-based Feature Encoder}
We developed a 3D convolutional neural network specifically designed to extract meaningful features from the 3D sMRI data. Without loss of generality, we utilize the source encoder $\mathcal{E}_{\text{s}}$ with a source image $\x_{\text{s}}$ as an example (see Fig.~\ref{overall_framework}). The source encoder $\mathcal{E}_{\text{s}}$ consists of 10 convolutional layers, each equipped with $3\times 3\times 3$ kernels to capture intricate spatial patterns in 3D brain images. Each convolutional layer is followed by batch normalization and $\operatorname{ReLU}$ activation, and downsampling is strategically applied to the even-numbered convolutional layers to enable hierarchical feature extraction. Recognizing the importance of specific brain regions in diagnosing various neurological disorders, as highlighted by previous studies \cite{lian2018hierarchical, mu2011adult, woo2018cbam}, we integrated an attention mechanism within our network. For this purpose, the output feature maps from the final layer of $\mathcal{E}_{\text{s}}$ are fed into the spatial attention module $\mathcal{M}(\cdot)$, where the location-based global attention is applied. Specifically, the output feature maps $\mathbf{F}_{\text{s}}=\mathcal{E}_{\text{s}}(\x_{\text{s}})$ undergo respective max-pooling $\operatorname{MP}(\cdot)$ and average-pooling $\operatorname{AP}(\cdot)$, which are then concatenated to integrate the information from these different perspectives. The spatial attention module finally refines these merged features by a convolutional operation, followed by Sigmoid activation $\sigma$ to quantify the attentive scores as
\begin{equation}
    \mathbf{S}_{\text{s}} = \sigma\left(\operatorname{Conv1D}\left(\operatorname{MP}(\mathbf{F}_{\text{s}})\oplus\operatorname{AP}(\mathbf{F}_{\text{s}})\right)\right),
\end{equation}

where $\oplus$ denotes the channel-wise concatenation. By multiplying the spatial attention map $\mathbf{S}_{\text{s}}$ by the output feature maps $\mathbf{F}_{\text{s}}$, we generate the spatial attentive features $\mathbf{F}'_{\text{s}}$ that highlight the most prominent areas regarding brain disease identification:
\begin{equation}
    \mathbf{F}'_{\text{s}} = \mathcal{M}(\mathbf{F}_{\text{s}}) = \mathbf{F}_{\text{s}} \odot \mathbf{S}_{\text{s}},
\end{equation}
where $\odot$ denotes the Hadamard product operation. Such spatial attentive features $\mathbf{F}'_{\text{s}}$ are fed into intensity discriminator $\mathcal{C}_{\text{I}}$ and label classifier $\mathcal{C}_{\text{L}}$ to differentiate the intensity variations across domains and to accurately predict the corresponding disease labels, respectively.

\subsubsection{Objective Functions}
To enforce the model's resilience to variations in image intensity, we employ a gradient reversal layer \cite{ganin2016domain}, which effectively inverts the gradient during backpropagation through self-adversarial learning, defined as

\begin{equation}\label{int_loss}
    \mathcal{L}_{\text{int}} = \operatorname{CE}(\mathcal{C}_{\text{I}}(\mathbf{F}), \mathbf{Y}_{\text{s}}),
\end{equation}
where $\y_{\text{s}}$ indicates the source category label and $\mathbf{F} \in \{\mathbf{F}'_{\text{s}}, \hat{\mathbf{F}}'_{\text{s}}\}$, with $\mathbf{F}'_{\text{s}},\hat{\mathbf{F}}'_{\text{s}}=\mathcal{M}(\mathcal{E}_{\text{s}}(\x_{\text{s}})),\mathcal{M}(\mathcal{E}_{\text{s}}(\hat{\x}_{\text{s}}))$, respectively. This learning strategy allows the model to learn features that are invariant to intensity-based texture changes. Additionally, the cross-entropy (CE) loss function is applied to enhance the model's capabilities for disease identification:
\begin{equation}\label{cls_s_loss}
    \mathcal{L}_{\text{cls}} = \operatorname{CE}\left(\mathcal{C}_{\text{L}}\left(\mathbf{F}\right), \mathbf{Y}_{\text{s}}\right).
\end{equation}
Accordingly, the complete objective function for the pertaining stage is defined as follows:
\begin{equation}
    \mathcal{L}_{\text{total}_1} = \mathcal{L}_{\text{cls}} - \mathcal{L}_{\text{int}}.
\end{equation}

\subsection{UDA via Dynamic Frequency Manipulation}
Following the pretraining step, the target encoder $\mathcal{E}_{\text{t}}$ for UDA is trained by applying the proposed DyMix for data manipulation to properly adapt the source and target domains while leveraging the knowledge transferred from the pretrained source encoder. As a preliminary step, the target encoder $\mathcal{E}_{\text{t}}$ is initialized by replicating both the architecture and the pretrained weights of the source encoder $\mathcal{E}_{\text{s}}$ before commencing the UDA process. This approach ensures that the target encoder benefits from the robust feature representations learned by the source encoder, providing a strong foundation for effective domain adaptation.

\subsubsection{DyMix Strategy}
Inspired by the mixup \cite{zhang2017mixup} technique within the frequency space, DyMix engages in an effective image transformation that involves linear interpolation between the source and target amplitude components. DyMix is analogous to the standard mixup technique but contains a primary distinction: the region size for amplitude mixing is dynamically optimized by the tunable $\beta$ scheduler with step temperature $\tau$ during training in contrast to using the manually pre-defined manipulation region, which is one of the major drawbacks of the conventional mixup \cite{zhang2017mixup} technique. 

The process begins by determining the initial region size, which is a critical aspect of our proposed DyMix. If the magnitude of the mixing region is not specified at the outset, it is rigid to the maximum possible size ($\beta=1$) for broad exploration (\ie, full-scale amplitude mixup). The $\beta$ scheduler then enters a conditional loop-continuously comparing the latest evaluation score to the best score recorded so far, and only holding on to a $\beta$ magnitude if the performance on a held-out validation set improves. Conversely, when there is no performance gain until the defined patience condition, the DyMix infers that the current region size may be limited to further progress. At this point, the DyMix performs a validation step by adjusting the $\beta$ using a step temperature $\tau$ that either increases or decreases the $\beta$ magnitude (\ie, $\beta_+=\beta+\tau$ or $\beta_-=\beta-\tau$). The region size based on the modified $\bar{\beta}$ that yields the highest evaluation score during validation is selected for the next round of training as
\begin{equation}\label{eq:amp-trans}
    \bar{\beta}=\left\{\begin{array}{ll}
    \beta + \tau & \text { if } Eval(\beta_+) > Eval(\beta_-)\\
    \beta - \tau & \text { otherwise, }
    \end{array}\right.
\end{equation}
where $Eval(\beta_+)$ and $Eval(\beta_-)$ respectively represent the validation performance when utilizing images that are manipulated by $\beta_+$ and $\beta_-$ magnitude-based DyMix, respectively. To further maintain the stable exploration for the amplitude mixup adjustment, constraints are applied to ensure the region size remains within the specified minimum and maximum. Algorithm~\ref{alg:dymix} describes the details of the implementation steps for DyMix.

\begin{algorithm}[!t]
\footnotesize
\caption{Pseudo algorithm for Dynamic Frequency Mixup Scheduler (DyMix)}
\label{alg:dymix}
\begin{algorithmic}[1]
\Require{Initial region size $\beta$, Step temperature $\tau$, Classification model $net$, Initial hyper-parameter settings : $best\_score \gets 0$, $num\_bad\_epochs \gets 0$, $patience \gets 5$, $max\_region \gets 1.0$, $min\_region \gets 0.0$, $\beta \gets 1.0$}

\State {\em // Validate the model during the training process}
\State {\em // Input $auc\_score$ through model validation}

\Function{Step}{$auc\_score$}
    \If{$auc\_score > best\_auc$}
        \State Update the $best\_auc \gets val\_score$
        \State Reset $num\_bad\_epochs \gets 0$
    \Else
        \State Increment $num\_bad\_epochs \gets num\_bad\_epochs + 1$
        \If{$num\_bad\_epochs > patience$}
            \State Adjust the region size $\beta$ using a step temperature $\tau$:
            \State $\beta_+, \beta_- \gets \beta+\tau, \beta-\tau$
            \State Evaluate the model $net$ on $\beta_+$ and $\beta_-$ based DyMix:
            \If{$Eval(\text{DyMix}(\beta_+)) > Eval(\text{DyMix}(\beta_-))$}
                \State Update the $\bar{\beta} \gets \beta_+$ 
            \Else
                \State Update the $\bar{\beta} \gets \beta_-$
            \EndIf
            \State Reset the $num\_bad\_epochs \gets 0$        
        \EndIf
    \EndIf
    \State \textbf{return} region\_size ($\bar{\beta} \times \bar{\beta}$)
\EndFunction
\end{algorithmic}
\end{algorithm}

\subsubsection{DyMix-based Data Manipulation}
For the application of DyMix, the source $\x_{\text{s}}$ and target $\x_{\text{t}}$ domain data are first transformed to the frequency spectrum through Eqs. \eqref{eq:amp} and \eqref{eq:pha} to obtain the amplitude $\mathcal{A}(\x_{\text{s}}), \mathcal{A}(\x_{\text{t}})$ and phase $\mathcal{P}(\x_{\text{s}}), \mathcal{P}(\x_{\text{t}})$ components, respectively. Following the FFT operation, the DyMix-based manipulation is conducted according to the tunable $\beta$ scheduler to blend the specific region between the source and target amplitude components as
\begin{equation}\label{amp_mix}
    \mathcal{A}_{\text{mix}} = (1-\lambda)\cdot\mathcal{A}_{\beta\times\beta}(\x_{\text{s}}) + \lambda\cdot\mathcal{A}_{\beta\times\beta}(\x_{\text{t}}),
\end{equation}
where $\lambda \sim U(0,1)$ refers to a random value drawn from a uniform distribution over the range $[0,1]$. The resulting mixed amplitude component $\mathcal{A}_{\text{mix}}$ is then combined with the phase component of the target image $\mathcal{P}(\x_{\text{t}})$ and processed through the iFFT $\mathcal{F}^{-1}$ to create the reconstructed amplitude-mixed target image $\hat{\x}_{\text{t}}$ as follows:
\begin{equation}\label{dmt}
    \hat{\x}_{\text{t}} = \mathcal{F}^{-1}(\mathcal{A}_{\text{mix}},\mathcal{P}(\x_{\text{t}})).
\end{equation}

\subsubsection{Objective Functions}
To strengthen the robustness of our model regardless of domain differences, it is crucial to maintain consistency between the attention maps of the source and target domains, denoted as $\mathbf{F}'_{\text{s}}=\mathcal{M}(\mathcal{E}_{\text{s}}(\x_{\text{s}}))$ and $\hat{\mathbf{F}}'_{\text{t}}=\mathcal{M}(\mathcal{E}_{\text{t}}(\hat{\x}_{\text{t}}))$, respectively. To achieve this, we introduce an attention consistency loss using $\ell$-2 regularization designed to facilitate the seamless imposition of semantically highlighted characteristics from the source data to the target data:
\begin{equation}\label{att_loss}
    \mathcal{L}_{\text{att}} = \frac{1}{H\times W\times D}\sum^{H}_{h=1}\sum^{W}_{w=1}\sum^{D}_{d=1}\left\Vert \mathbf{F}'_{\text{s}} - \hat{\mathbf{F}}'_{\text{t}}\right\Vert_2.
\end{equation}
In doing so, regularizing attention consistency ensures that the model mutually pays attention to the brain disease-associated location between spatial attentive features $\mathbf{F}'_{\text{s}}$ and $\hat{\mathbf{F}}'_{\text{t}}$.

In terms of domain knowledge distillation, the pretrained source encoder $\mathcal{E}_{\text{s}}$ and the label classifier $\mathcal{C}_{\text{L}}$ are used to assist the target encoder $\mathcal{E}_{\text{t}}$ in learning robust and meaningful feature representations for disease identification:
\begin{equation}\label{cls_t_loss}
    \mathcal{L}_{\text{cls}} = \operatorname{CE}(\mathcal{C}_{\text{L}}(\mathbf{F}'_{\text{s}}), \mathbf{Y}_{\text{s}}),
\end{equation}
where $\y_{\text{s}}$ denotes the source category label.

To further mitigate the domain discrepancy between the source and target domains, we implemented a domain discriminator $\mathcal{C}_{\text{D}}$ using cross-entropy (CE) loss, which functions similarly to the training of the intensity discriminator $\mathcal{C}_{\text{I}}$ in the pretraining step. By doing so, $\mathcal{C}_{\text{D}}$ performs to differentiate between brain features originating from different domains, effectively identifying domain-specific characteristics:
\begin{equation}\label{dom_loss}
    \mathcal{L}_{\text{dom}} = \operatorname{CE}(\mathcal{C}_{\text{D}}(\mathbf{F}), \mathbf{Y}_{\text{d}}),
\end{equation}
where $\y_{\text{d}}\in\{0,1\}$ indicates the domain label and $\mathbf{F}\in\{\mathbf{F}'_{\text{s}}, \hat{\mathbf{F}}'_{\text{t}}\}$ with $\mathbf{F}'_{\text{s}},\hat{\mathbf{F}}'_{\text{t}}=\mathcal{M}(\mathcal{E}_{\text{s}}(\x_{\text{s}})),\mathcal{M}(\mathcal{E}_{\text{t}}(\hat{\x}_{\text{t}}))$, respectively.

The overall objective function is structured for minimization, despite including a term that involves maximizing the domain discrimination loss $\mathcal{L}_{\text{dom}}$. This is entailed using a gradient reversal layer in the domain discriminator $\mathcal{C}_{\text{D}}$, which inverts the gradients during the backward process by multiplying by a negative constant, thereby maximizing $\mathcal{L}_{\text{dom}}$. In this light, our training strategy involves simultaneously minimizing the label classification loss $\mathcal{L}_{\text{cls}}$ and the attention consistency loss $\mathcal{L}_{\text{att}}$, while maximizing the domain discrimination loss $\mathcal{L}_{\text{dom}}$:
\begin{equation}\label{total_loss}
    \mathcal{L}_{\text{total}_2} = \mathcal{L}_{\text{cls}} + \lambda_1\mathcal{L}_{\text{att}} - \lambda_2\mathcal{L}_{\text{dom}},
\end{equation}
where $\lambda_1$ and $\lambda_2$ are the weight coefficients used to balance the contribution of  $\mathcal{L}_{\text{att}}$, $\mathcal{L}_{\text{dom}}$ loss function.

\begin{table*}[t!]\centering\scriptsize\setlength{\tabcolsep}{6.5pt}
    \caption{Performance metrics (\%) of the proposed method compared to various UDA baselines in AD diagnosis across different domain transfer settings (including AD $vs.$ CN, MCI $vs.$ AD, and CN $vs.$ MCI scenarios). Here, the abbreviations of ACC, SEN, SPE, and AUC denote the accuracy, sensitivity, specificity, and area under the ROC curve, respectively.}\label{tab_DA}
    \begin{tabular}{lccccc|cccc|cccc}
        \toprule
        \multicolumn{1}{c}{\textbf{Domain Transfer Settings}} & \multicolumn{1}{c}{\multirow{2}{*}{\textbf{Methods}}} & \multicolumn{4}{c}{\textbf{AD $vs.$ CN Scenario}} & \multicolumn{4}{c}{\textbf{MCI $vs.$ AD Scenario}} & \multicolumn{4}{c}{\textbf{CN $vs.$ MCI Scenario}}\\
        \cmidrule(lr){3-6} \cmidrule(lr){7-10} \cmidrule(lr){11-14}
        \multicolumn{1}{c}{(\textbf{Source $\rightarrow$ Target})} & & \textbf{ACC} & \textbf{SEN} & \textbf{SPE} & \textbf{AUC} & \textbf{ACC} & \textbf{SEN} & \textbf{SPE} & \textbf{AUC} & \textbf{ACC} & \textbf{SEN} & \textbf{SPE} & \textbf{AUC}\\
        \midrule
        \multirow{8}{*}{ADNI-1 $\rightarrow$ \text{ADNI-2}} & Source-Only & 83.33 & 65.62 & 97.50 & 81.56 & 52.68 & 55.08 & 36.73 & 50.91 & 52.68 & 55.08 & 36.73 & 50.91 \\
        & Target-Only & 98.61 & 96.87 & 100.0 & 98.43 & 75.00 & 78.25 & 83.78 & 78.56 & 65.18 & 68.25 & 61.22 & 64.74 \\ \cmidrule(l){2-14}
        & DANN \cite{ganin2016domain} & 84.77 & 75.00 & 92.50 & 83.75 & 53.57 & 58.73 & 46.94 & 52.83 & 53.57 & 58.73 & 46.94 & 52.83 \\
        & Deep-CORAL \cite{sun2017correlation} & 84.72 & 75.00 & 92.50 & 83.75 & 56.25 & 64.13 & 40.41 & 52.27 & 56.25 & 54.13 & 20.41 & 52.27 \\
        & AD$^{2}$A \cite{guan2021multi} & 86.11 & 75.00 & 95.00 & 85.00 & 56.25 & 60.32 & 51.02 & 55.67 & 56.25 & 60.32 & 51.02 & 55.67 \\
        & PMDA \cite{cai2023prototype} & 90.28 & 81.25 & 97.50 & 89.37 & 58.27 & 43.33 & 51.08 & 52.21 & 55.36 & 55.24 & 47.96 & 54.57 \\
        & FMM \cite{shin2023frequency} & 90.28 & 81.25 & 97.50 & 89.37 & 59.82 & 66.67 & 51.02 & 58.84 & 59.82 & 66.67 & 51.02 & 58.84 \\
        & \cellcolor{mygray}\textbf{Ours} & \cellcolor{mygray}\textbf{91.67} & \cellcolor{mygray}\textbf{81.25} & \cellcolor{mygray}\textbf{100.0} & \cellcolor{mygray}\textbf{90.62} & \cellcolor{mygray}\textbf{71.15} & \cellcolor{mygray}\textbf{73.33} & \cellcolor{mygray}\textbf{70.27} & \cellcolor{mygray}\textbf{71.80}& \cellcolor{mygray}\textbf{61.78} & \cellcolor{mygray}\textbf{67.14} & \cellcolor{mygray}\textbf{54.90} & \cellcolor{mygray}\textbf{61.02} \\
        \midrule
        \multirow{8}{*}{\text{ADNI-1+ADNI-2} $\rightarrow$ \text{ADNI-3}} & Source-Only & 80.00 & 82.31 & 79.55 & 80.93 & 55.00 & 36.36 & 45.36 & 60.86 & 52.38 & 66.47 & 40.84 & 58.66 \\ 
        & Target-Only & 98.75 & 92.31 & 100.0 & 96.15 & 86.54 & 83.64 & 92.68 & 88.16 & 79.52 & 78.23 & 94.51 & 81.37 \\ \cmidrule(l){2-14}
        & DANN \cite{ganin2016domain} & 81.25 & 84.62 & 74.78 & 84.69 & 46.15 & 50.00 & 31.71 & 65.85 & 58.09 & 73.53 & 50.70 & 62.12 \\
        & Deep-CORAL \cite{sun2017correlation} & 85.00 & 92.31 & 83.58 & 87.94 & 58.85 & 48.18 & 50.00 & 59.09 & 65.71 & 55.88 & 70.42 & 63.15 \\
        & AD$^{2}$A \cite{guan2021multi} & 90.00 & \textbf{100.0} & 88.06 & 90.03 & 65.38 & 61.82 & 60.97 & 71.40 & 60.48 & 70.59 & 60.84 & 55.72 \\
        & PMDA \cite{cai2023prototype} & 83.75 & 84.62 & 88.06 & 91.86 & 66.54 & 54.55 & \textbf{95.12} & 74.83 & 59.05 & 74.12 & 62.68 & 63.40 \\
        & FMM \cite{shin2023frequency} & 88.75 & 92.31 & 88.06 & 93.14 & 78.85 & 63.64 & 82.93 & 73.28 & 63.33 & 77.06 & 66.34 & 71.70 \\
        & \cellcolor{mygray}\textbf{Ours} & \cellcolor{mygray}\textbf{91.25} & \cellcolor{mygray}92.31 & \cellcolor{mygray}\textbf{89.10} & \cellcolor{mygray}\textbf{95.70} & \cellcolor{mygray}\textbf{80.77} & \cellcolor{mygray}\textbf{72.73} & \cellcolor{mygray}82.93 & \cellcolor{mygray}\textbf{77.83} & \cellcolor{mygray}\textbf{70.25} & \cellcolor{mygray}\textbf{79.41} & \cellcolor{mygray}\textbf{75.21} & \cellcolor{mygray}\textbf{77.31} \\
        \midrule
        \multirow{8}{*}{\text{ADNI-1} $\rightarrow$ \text{AIBL}} & Source-Only & 77.93 & 72.61 & 79.25 & 75.93 & 56.67 & 20.00 & 75.83 & 57.92 & 54.92 & 55.00 & 50.00 & 62.50 \\ 
        & Target-Only & 97.93 & 76.52 & 95.70 & 86.11 & 79.23 & 76.67 & 95.83 & 81.25 & 83.61 & 80.83 & 98.98 & 79.91 \\ \cmidrule(l){2-14}
        & DANN \cite{ganin2016domain} & 74.14 & 69.56 & 75.27 & 72.42 & 58.97 & 53.33 & 50.00 & 61.67 & 78.69 & 58.33 & 83.67 & 71.00 \\
        & Deep-CORAL \cite{sun2017correlation} & 87.07 & 65.22 & 88.17 & 80.46 & 64.10 & 46.67 & 75.00 & 60.83 & 77.87 & 59.17 & 89.79 & 69.48 \\
        & AD$^{2}$A \cite{guan2021multi} & 85.34 & 73.91 & 88.17 & 81.04 & 71.79 & 60.00 & 79.17 & 69.58 & 72.29 & 64.17 & 84.28 & 69.23 \\
        & PMDA \cite{cai2023prototype} & 86.21 & 60.87 & 92.47 & 76.67 & 71.54 & 60.00 & 75.00 & 67.50 & 77.87 & 70.83 & 87.35 & 74.09 \\
        & FMM \cite{shin2023frequency} & 89.65 & 78.26 & 92.47 & 85.37 & 76.92 & 66.67 & 83.33 & 75.00 & 80.10 & 72.50 & 84.08 & \textbf{78.84} \\
        & \cellcolor{mygray}\textbf{Ours} & \cellcolor{mygray}\textbf{91.38} & \cellcolor{mygray}\textbf{78.27} & \cellcolor{mygray}\textbf{94.62} & \cellcolor{mygray}\textbf{86.44} & \cellcolor{mygray}\textbf{78.97} & \cellcolor{mygray}\textbf{66.67} & \cellcolor{mygray}\textbf{86.67} & \cellcolor{mygray}\textbf{76.67}& \cellcolor{mygray}\textbf{81.57} & \cellcolor{mygray}\textbf{79.17} & \cellcolor{mygray}\textbf{92.04} & \cellcolor{mygray}77.69 \\
        \midrule 
        \multirow{8}{*}{\text{ADNI-1+ADNI-2} $\rightarrow$ \text{AIBL}} & Source-Only & 77.07 & 73.91 & 80.32 & 72.11 & 59.23 & 53.33 & 49.17 & 56.25 & 43.44 & 40.83 & 36.73 & 43.78 \\ 
        & Target-Only & 97.93 & 76.52 & 95.70 & 86.11 & 79.23 & 76.67 & 95.83 & 81.25 & 83.61 & 80.83 & 98.98 & 79.91 \\ \cmidrule(l){2-14}
        & DANN \cite{ganin2016domain} & 81.55 & 81.30 & 86.67 & 78.98 & 64.10 & 63.33 & 50.00 & 61.67 & 50.82 & 41.84 & 87.50 & 64.67 \\
        & Deep-CORAL \cite{sun2017correlation} & 80.52 & 60.87 & 87.85 & 79.36 & 58.97 & 66.67 & 54.17 & 60.42 & 72.95 & 55.00 & 84.69 & 54.85 \\
        & AD$^{2}$A \cite{guan2021multi} & 85.34 & 73.91 & 88.17 & 80.04 & 66.67 & 53.33 & 75.00 & 64.17 & 69.02 & 62.50 & 88.16 & 60.33 \\
        & PMDA \cite{cai2023prototype} & 80.17 & 78.26 & 81.72 & 79.45 & 69.23 & 56.67 & 75.83 & 61.25 & 75.41 & 54.17 & 91.86 & 58.51 \\
        & FMM \cite{shin2023frequency} & 86.21 & 79.56 & 90.32 & 79.94 & 71.79 & 66.67 & 75.00 & 70.83 & 74.75 & 55.83 & 89.39 & 67.61 \\
        & \cellcolor{mygray}\textbf{Ours} & \cellcolor{mygray}\textbf{89.31} & \cellcolor{mygray}\textbf{82.61} & \cellcolor{mygray}\textbf{90.65} & \cellcolor{mygray}\textbf{80.55} & \cellcolor{mygray}\textbf{74.36} & \cellcolor{mygray}\textbf{66.67} & \cellcolor{mygray}\textbf{79.17} & \cellcolor{mygray}\textbf{72.92}& \cellcolor{mygray}\textbf{81.57} & \cellcolor{mygray}\textbf{62.50} & \cellcolor{mygray}\textbf{92.04} & \cellcolor{mygray}\textbf{67.68} \\
        \midrule
        \multirow{8}{*}{\text{AIBL} $\rightarrow$ \text{ADNI-3}} & Source-Only & 80.00 & 36.15 & 88.51 & 72.33 & 76.54 & 33.64 & 82.68 & 58.16 & 57.62 & 35.88 & 77.18 & 41.53 \\
        & Target-Only & 98.75 & 92.31 & 100.0 & 96.15 & 86.54 & 83.64 & 92.68 & 88.16 & 79.52 & 78.23 & 94.51 & 81.37 \\ \cmidrule(l){2-14}
        & DANN \cite{ganin2016domain} & 88.75 & 40.77 & 90.00 & 75.38 & 84.61 & 36.36 & 97.56 & 66.96 & 62.86 & 44.71 & 85.92 & 50.31 \\
        & Deep-CORAL \cite{sun2017correlation} & 86.25 & 53.85 & 92.54 & 73.19 & 80.77 & 29.09 & \textbf{100.0} & 54.55 & 61.90 & 41.76 & 85.92 & 48.84 \\
        & AD$^{2}$A \cite{guan2021multi} & 81.25 & \textbf{56.92} & 82.09 & 79.51 & 80.77 & 45.45 & 90.24 & 67.85 & 69.52 & 44.71 & 95.77 & 55.24 \\
        & PMDA \cite{cai2023prototype} & 83.75 & 30.77 & 85.07 & 77.92 & 80.77 & 27.27 & 92.68 & 59.98 & 66.67 & 40.00 & \textbf{98.59} & 49.30 \\
        & FMM \cite{shin2023frequency} & 90.00 & 53.85 & 97.01 & 75.43 & 81.76 & 72.73 & 82.93 & 77.83 & 65.71 & 59.41 & 93.10 & 56.25 \\
        & \cellcolor{mygray}\textbf{Ours} & \cellcolor{mygray}\textbf{91.25} & \cellcolor{mygray}53.85 & \cellcolor{mygray}\textbf{98.51} & \cellcolor{mygray}\textbf{80.18} & \cellcolor{mygray}\textbf{84.62} & \cellcolor{mygray}\textbf{81.82} & \cellcolor{mygray}85.36 & \cellcolor{mygray}\textbf{83.59}& \cellcolor{mygray}\textbf{70.66} & \cellcolor{mygray}\textbf{68.82} & \cellcolor{mygray}93.37 & \cellcolor{mygray}\textbf{61.59} \\
        \bottomrule
    \end{tabular}
\end{table*}

\section{Experiments}
\subsection{Datasets and Data Preprocessing}
To demonstrate the validity of our study, we utilized two publicly available benchmark datasets: the Alzheimer’s Disease Neuroimaging Initiative (ADNI)~\cite{mueller2005alzheimer} and the Australian Imaging Biomarker and Lifestyle Study of Aging (AIBL)~\cite{rowe2010amyloid}. 

The ADNI dataset, a well-known and widely used resource in AD research, provides longitudinal data from individuals diagnosed with AD and MCI and cognitively normal (CN) individuals. It encompasses diverse demographic information, including clinical assessments, neuroimaging scans (MRI and positron emission tomography), genetic information, and biomarker measurements. The ADNI dataset comprises three sub-datasets, including ADNI1, ADNI2, and ADNI3. The three sub-datasets comprised 2,153 1.5T T1-weighted sMRI scans distributed as follows: the ADNI-1 dataset comprised 231 CN subjects, 414 subjects diagnosed with MCI, and 200 subjects diagnosed with AD; the ADNI-2 dataset comprised 201 CN subjects, 357 subjects diagnosed with MCI, and 159 subjects diagnosed with AD; and the ADNI-3 dataset comprised 332 CN subjects, 193 subjects diagnosed with MCI, and 66 subjects diagnosed with AD.

The AIBL dataset is a significant Australian research initiative designed to investigate early biomarkers and the underlying causes of AD. It also includes a comprehensive range of 
demographic information, similar to the ADNI. The AIBL comprised 689 subjects: 83 subjects diagnosed with AD, 112 subjects diagnosed with MCI, and 494 CN individuals.

In this work, the brain scans from both the ADNI and AIBL datasets were identically preprocessed using a defined pipeline. First, the HD-BET brain extraction tool~\cite{isensee2019automated} was employed to remove non-brain tissues from the MRI images, such as the neck and skull. The resulting skull-stripped images were then aligned to the MNI152 template using the FLIRT linear image registration tool from the FMRIB Software Library v6.0.1 (FSL)~\cite{zhang2001segmentation}. Afterward, these alignments corrected for global linear differences, including translation, scale, and rotation, and were normalized to a uniform spatial resolution (\ie, $1 \times 1 \times 1$ $mm^3$). Subsequently, the preprocessed 3D brain scans were obtained, each with a dimensionality of $193 \times 229 \times 193$. Finally, each image was finally normalized by using minmax normalization, scaling the voxel values to a range of $[0,1]$.

\begin{figure}[!t]
    \centering
    \includegraphics[width=\columnwidth]{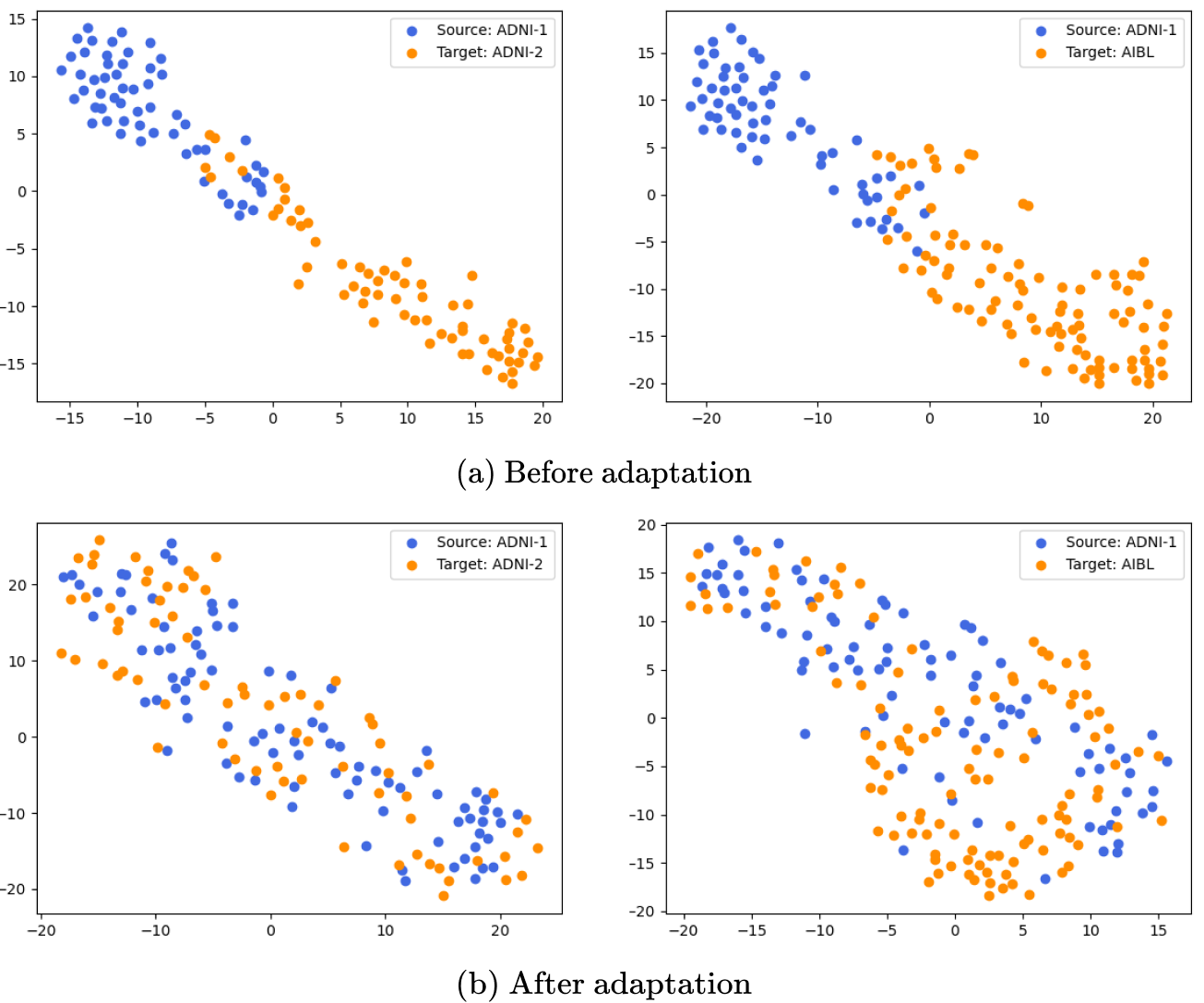}
    \caption{A t-SNE visualization of (a) the original distribution and (b) the distribution after domain adaptation using our proposed DyMix technique. These visualizations compared the source and target domains across different UDA scenarios, specifically ADNI-1 $\rightarrow$ ADNI-2 (first column) and a ADNI-1 $\rightarrow$ AIBL (second column).}
    \label{fig:tsne}
\end{figure}

\subsection{Experimental Setup}
Following various disease categories of these two benchmark datasets, we first established three scenarios for thorough experiments: (1) AD identification (\ie, AD $vs.$ CN scenario), (2) AD conversion identification (\ie, MCI $vs.$ AD scenario), and (3) early detection of cognitive decline (\ie, CN $vs.$ MCI scenario).

\subsubsection{Domain Transfer Settings}
To comprehensively assess the UDA ability of our proposed method, the ADNI dataset was carefully divided into three distinct sub-datasets: ADNI-1, ADNI-2, and ADNI-3. ADNI-1 and ADNI-2 are primarily focused on tracking the progression of AD through the analysis of various biological markers and changes in cognitive function over time. ADNI-3, being the most recent phase, builds on the findings of its predecessors with more advanced imaging techniques and refined biomarker measurements. From this perspective, while all three datasets aim to advance AD research, the participant pools and data collection protocols differ across these sub-datasets owing to the evolution of scientific knowledge and technological advancements over time. Accordingly, we treated ADNI-1 and ADNI-2 as distinct domains, wile we treated ADNI-3 as an additional domain for evaluating our UDA scenarios. In detail, we constructed two domain transfer settings (\ie, source domain $\rightarrow$ target domain) within the ADNI as (1) ADNI-1 $\rightarrow$ ADNI-2 and (2) ADNI-1+ADNI-2 $\rightarrow$ ADNI-3. To further validate the robustness of our approach across a completely different range of contexts, we also incorporated the AIBL dataset as another domain, allowing us to assess the generalizability and effectiveness of our method rigorously: (3) ADNI-1 $\rightarrow$ AIBL, (4) ADNI-1+ADNI-2 $\rightarrow$ AIBL, and reversed case (5) AIBL $\rightarrow$ ADNI-3.

\subsubsection{Implementation Details}
The proposed model was implemented in Python using the PyTorch framework. Two discriminators, including $\mathcal{C}_{\text{I}}$ and $\mathcal{C}_{\text{D}}$, and the label classifier $\mathcal{C}_{\text{L}}$ were identically constructed by three fully connected layers with 128, 64, and 2 units. The network was trained over 100 epochs with the Adam optimizer \cite{kingma2014adam}, set to a learning rate of 0.0001 and a batch size of 4. To prevent the risk of overfitting, we employed a dropout rate of 0.5 during the training. The hyperparameters $\lambda_1$ and $\lambda_2$ in Eq.~\eqref{total_loss} were empirically set to 0.5 and 0.1, respectively. 

\subsubsection{Evaluation Metrics} For the quantitative evaluation of our proposed method, we utilized four widely recognized criteria to assess classification performance: accuracy (ACC), sensitivity (SEN), specificity (SPE), and the area under the receiver operating characteristic (ROC) curve (AUC). These metrics provide an exhaustive understanding of the model’s effectiveness in distinguishing between different categories, and the higher the value for each metric, the better the performance of the model.

\subsubsection{Training Configurations}
During the second phase of model training, we initially pretrained the source encoder $\mathcal{E}_{\text{s}}$ and the attention module $\mathcal{M}$ for classification for 50 epochs using Eq.~\eqref{cls_s_loss}. Subsequently, these modules were fine-tuned and co-trained with both the domain discriminator and the category classifier in accordance with Eq.~\eqref{total_loss}. Meanwhile, the initial configuration for the DyMix was set with a step temperature $\tau$ of 0.05, a patience threshold of 5, a minimum amplitude region of 0.1, and a maximum amplitude region of 1.0. The optimal model configuration was selected based on the AUC score, employing a simple hold-out validation strategy to ensure both the robustness and reliability of the results. All experiments were executed on a workstation powered by an NVIDIA TITAN RTX GPU with 24GB of memory.

\subsection{Quantitative Results and Qualitative Analyses}
To conduct a comprehensive assessment, we compared the proposed method with several state-of-the-art UDA methods~\cite{sun2017correlation, ganin2016domain, guan2021multi} that are widely utilized in contemporary medical imaging tasks. For a fair comparison, we employed the architecture of our encoder $\mathcal{E}$ as the backbone feature extractor to implement and assess well-constructed UDA frameworks, such as DANN~\cite{ganin2016domain} and Deep-CORAL~\cite{sun2017correlation}. To establish the baseline performance benchmarks, we reported the results of the source-only (\ie, the lower bound) and target-only (\ie, the upper bound) methods, which were trained exclusively on a single domain without any adaptation. Table~\ref{tab_DA} summarizes the results of the baseline benchmarks and these evaluations.

In the AD $vs.$ CN scenario, our proposed method consistently outperformed other UDA methods across all domain transfer settings. A notable observation in this scenario is that the results of our DyMix in the ADNI-1 $\rightarrow$ AIBL setting achieved a substantial performance improvement of 6.89$\%$ in ACC and 7.25$\%$ in AUC compared to the average performance of the UDA methods. This highlights DyMix's superior capability to effectively regulate domain discrepancies and enhance model generalization. In the MCI $vs.$ AD scenario, our method derived the highest overall performance across all domain transfer settings, underscoring its robustness and adaptability to diverse domain shifts.

In the CN $vs.$ MCI scenario, particularly in the ADNI-1 $\rightarrow$ ADNI-2 and ADNI-1 $\rightarrow$ AIBL settings, the ACC and AUC scores of our method were nearly on par with those of the target-only method, which is considered the upper limit of UDA performance. Notably, in the ADNI-1 $\rightarrow$ AIBL setting, the performance gap between our method and the target-only model was merely $\pm$2.04$\%$ in ACC and $\pm$2.22$\%$ in AUC. This result is considerably remarkable, given the inherent complexity of domain adaptation between two vastly different data distributions. The challenge is further compounded by the subtle morphological variations between CN and MCI, making distinguishing between these categories particularly difficult. Despite these complexities, our proposed method exhibited outstanding performance, demonstrating an impressive gain compared to the average performance of other UDA methods.

Fig.~\ref{fig:tsne} presents a t-SNE visualization that illustrates the distribution of data points before and after applying our DyMix technique for domain adaptation. In both scenarios (\ie, ADNI-1 $\rightarrow$ ADNI-2 and ADNI-1 $\rightarrow$ AIBL settings), DyMix demonstrated its effectiveness by aligning the source and target distributions closer together in the feature representation space. Consequently, the improved overlap between domains after adaptation suggests that the model is more capable of robust cross-domain AD classification.

\begin{table}[t!]\centering\scriptsize\setlength{\tabcolsep}{6pt}
    \caption{Comparison of domain adaptation capabilities of pretrained models within the source domain using frequency manipulation-based self-adversarial learning ($\mathcal{L}_{\text{int}}$) and effectiveness of incorporating attention consistency loss ($\mathcal{L}_{\text{att}}$) in improving AD $vs.$ CN classification performance.}\label{tab_ab1}
    \begin{tabular}{cccccc}
        \toprule
        \text{Source} $\rightarrow$ \text{Target} & \textbf{Method} & \textbf{ACC} & \textbf{SEN} & \textbf{SPE} & \textbf{AUC}\\
        \midrule
        \multirow{3}{*}{\text{ADNI-1} $\rightarrow$ \text{ADNI-2}} & w/o $\mathcal{L}_{\text{int}}$ & 87.50 & 71.87 & 100.0 & 85.93 \\
        & w/o $\mathcal{L}_{\text{att}}$ & 87.50 & 81.25 & 92.50 & 86.87 \\
        & \textbf{Ours} & \textbf{91.67} & \textbf{81.25} & \textbf{100.0} & \textbf{90.62} \\
        \midrule
        \multirow{3}{*}{\text{ADNI-1+ADNI-2} $\rightarrow$ \text{ADNI-3}} & w/o $\mathcal{L}_{\text{int}}$ & 80.00 & 82.31 & 77.62 & 84.96 \\
        & w/o $\mathcal{L}_{\text{att}}$ & 85.00 & 92.31 & 83.58 & 87.94 \\
        & \textbf{Ours} & \textbf{91.25} & \textbf{92.31} & \textbf{89.10} & \textbf{95.70} \\
        \midrule
        \multirow{3}{*}{\text{ADNI-1} $\rightarrow$ \text{AIBL}} & w/o $\mathcal{L}_{\text{int}}$ & 85.00 & 75.00 & 82.09 & 81.04 \\
        & w/o $\mathcal{L}_{\text{att}}$ & 88.79 & 69.56 & 93.55 & 81.56 \\
        & \textbf{Ours} & \textbf{91.38} & \textbf{78.27} & \textbf{94.62} & \textbf{86.44} \\
        \midrule
        \multirow{3}{*}{\text{ADNI-1+ADNI-2} $\rightarrow$ \text{AIBL}} & w/o $\mathcal{L}_{\text{int}}$ & 85.34 & 73.91 & 88.17 & 80.04 \\
        & w/o $\mathcal{L}_{\text{att}}$ & 82.50 & 82.61 & 79.10 & 79.48 \\
        & \textbf{Ours} & \textbf{89.31} & \textbf{82.61} & \textbf{90.65} & \textbf{80.55} \\
        \midrule
        \multirow{3}{*}{\text{AIBL} $\rightarrow$ \text{ADNI-3}} & w/o $\mathcal{L}_{\text{int}}$ & 81.90 & 49.56 & 84.95 & 77.25 \\
        & w/o $\mathcal{L}_{\text{att}}$ & 88.23 & 52.61 & 96.34 & 79.48 \\
        & \textbf{Ours} & \textbf{91.25} & \textbf{53.85} & \textbf{98.51} & \textbf{80.18} \\
        \bottomrule
    \end{tabular}
\end{table}

\subsection{Ablation Study} 
To thoroughly validate our proposed method, we conducted a series of ablation studies focusing on two critical aspects: (i) the robustness of invariant feature representations (without $\mathcal{L}_{\text{int}}$ during the first step training) and (ii) the effectiveness of the spatial attention module (without $\mathcal{L}_{\text{att}}$ during the second step training), as reported in Table~\ref{tab_ab1}. Through these analyses, we explored each component's unique contributions to the model's overall performance in UDA tasks for AD diagnosis and provided insights into their necessity.

\begin{figure}[!t]
    \centering
    \includegraphics[width=\columnwidth]{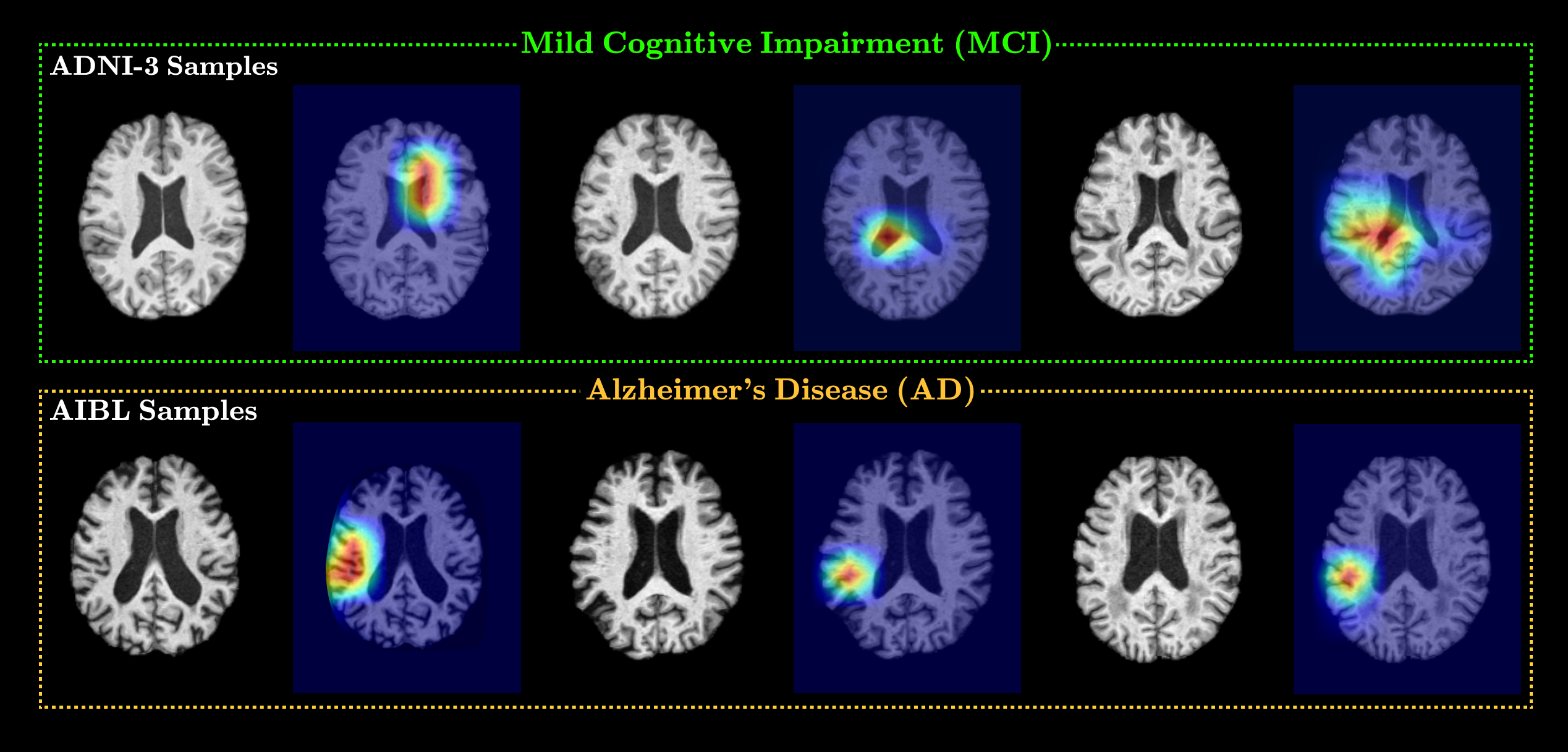}
    \caption{Illustration of model interpretability using Grad-CAM in the unseen target domain. Here, a red-colored region and a purple-colored region indicate a greater and lesser impact on the model decision, respectively.}
    \label{gradcam_visualization}
\end{figure}

\subsubsection{Robustness of Invariant Feature Representations}
The rationale behind UDA tasks, particularly in medical imaging, is that domain shifts due to differences in scanner protocols and intensity variations can drastically influence the model performance. For robust domain adaptation, it is crucial that the model learns feature representations that are invariant to these variations. From this perspective, we employed frequency manipulation to create intensity-transformed images and train the model using self-adversarial learning via $\mathcal{L}_{\text{int}}$.

By excluding $\mathcal{L}_{\text{int}}$ from the pretraining phase (\ie, w/o $\mathcal{L}_{\text{int}}$), we observed a considerable decline in performance across various domain adaptation scenarios. Notably, the ADNI-1+ADNI-2 $\rightarrow$ ADNI-3 transfer setting, where the source and target domains possess relatively similar data characteristics, showed a marked severe decrease in both ACC and AUC scores by -11.25$\%$ and -10.74$\%$, respectively. This indicates that even when the distribution discrepancy is relatively minor, the difficulty of invariant feature extraction under diverse intensity variations can degrade model performance. Furthermore, the performance drop was even more pronounced in more challenging settings, such as AIBL $\rightarrow$ ADNI-3, where the source and target domains represent entirely different datasets with distinct imaging properties. This observation further argues for the critical role of intensity-invariant feature learning in domain adaptation in complex cross-dataset environments. As a result, these findings demonstrated that manipulating the frequency domain to generate intensity-robust feature representations significantly enhances the model's ability to generalize across different domains.

\subsubsection{Effectiveness of Spatial Attention Mechanism}
The primary of the spatial attention mechanism with attention consistency loss $\mathcal{L}_{\text{att}}$ is to guide the model in consistently highlighting the most discriminative regions relevant to AD across different domains. Table~\ref{tab_ab1} presents the performance of these configurations across various domain adaptation settings. Compared to the results without $\mathcal{L}_{\text{att}}$ (\ie, w/o $\mathcal{L}_{\text{att}}$), our method clearly showed that incorporating $\mathcal{L}_{\text{att}}$ substantially improved the model's performance in all evaluation metrics. It means enforcing attention consistency helps the model focus on anatomically meaningful regions critical for distinguishing between disease states, enhancing its capacity to deliver robust diagnostic outcomes regardless of domain-specific variations.

To provide further insights, we exhibited the results of Grad-CAM~\cite{selvaraju2020grad} derived from the ADNI-3 and AIBL, as illustrated in Fig.~\ref{gradcam_visualization}. These saliency maps revealed that the most discriminative regions (red-colored regions), essential for AD prognosis, are primarily located in the ventricle, middle temporal gyrus, and superior temporal gyrus. Intriguingly, those discovered regions are well-recognized as key landmarks in AD progression~\cite{risacher2009baseline,davies1987quantitative}, particularly in the context of neurodegeneration. Based on such qualitative inspection, we are convinced that our spatial attention consistently focused on these critical regions across different domains.

\begin{figure}[!t]
    \centering
    \includegraphics[width=0.97\columnwidth]{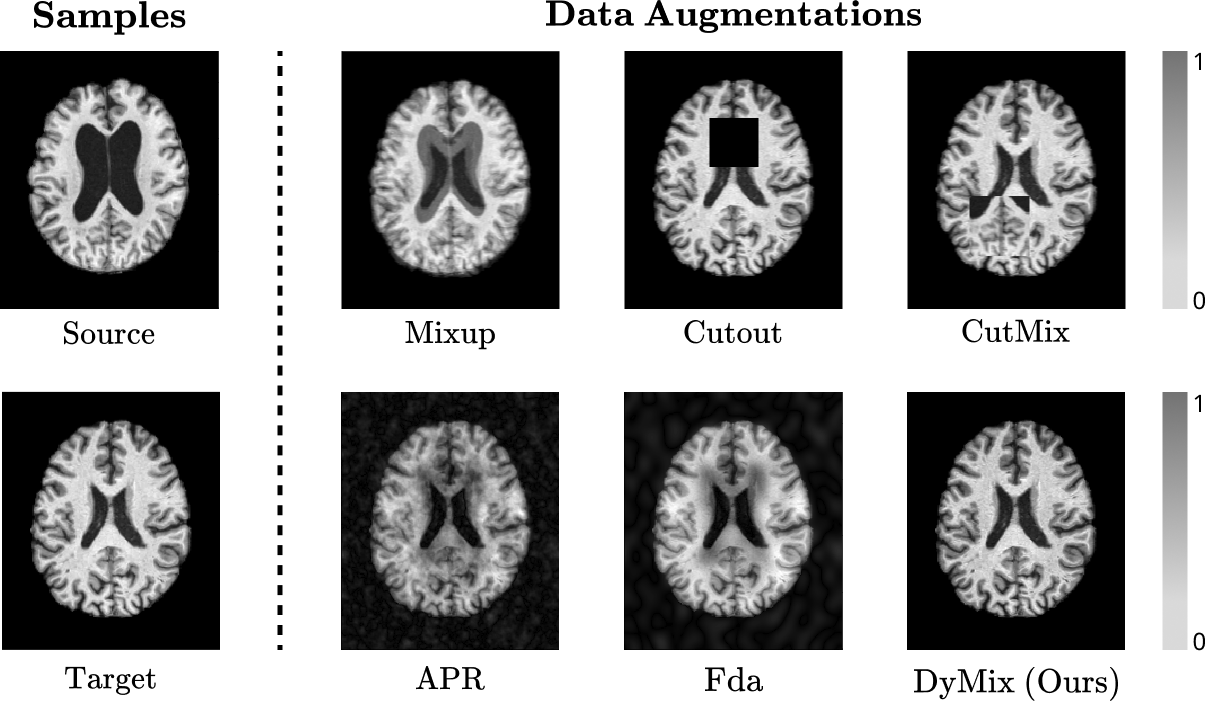}
    \caption{Visual examples of various spatial-based (top right) and frequency-based (bottom right) augmentation methods applied to both source and target brain MRI samples. Each augmentation highlights its unique effect on the image features, illustrating differences in how spatial and frequency components are manipulated to facilitate domain adaptation.}
    \label{fig:various_aug}
\end{figure}

\subsection{DyMix versus Various Augmentation Methods}
To assess the effectiveness of our proposed DyMix in comparison to existing data augmentation techniques, we conducted a series of experiments analyzing the impact of different augmentations on image quality and model performance in the AD $vs.$ CN scenario. In this regard, prevalent techniques of five data augmentations were adopted, including spatial-based (\ie, Mixup~\cite{zhang2017mixup}, Cutout~\cite{devries2017improved}, and CutMix~\cite{yun2019cutmix}) and frequency-based (\ie, APR~\cite{chen2021amplitude} and Fda~\cite{yang2020fda}) methods:
\begin{itemize}
    \item Mixup: creates new training samples by linearly interpolating pairs of examples, thereby smoothing the decision boundary between classes.
    \item Cutout: randomly masks out square regions of the input image, forcing the model to focus on less evident features.
    \item CutMix: combines two images by cutting and pasting patches between them, which enhances the model's ability to generalize by introducing more training samples.
    \item APR: recombines amplitude and phase information from different domains to enhance domain-invariant features.
    \item Fda: aligns source and target domains by swapping low-frequency components to smooth domain shifts.
\end{itemize}

Our proposed DyMix retained a high level of anatomical fidelity with a more balanced and context-aware transformation compared to other augmentation techniques, as illustrated in Fig.~\ref{fig:various_aug}. The brain structures remained clear and undistorted, preserving critical diagnostic features (\eg, the periventricular area) needed for accurate AD classification. In contrast, methods such as the conventional mixup and cutout techniques compromise image quality or overlook critical neuroanatomical details, which may result in suboptimal performance owing to the loss of essential morphological information (\ie, brain atrophies). Table~\ref{tab_ab3} confirms that our DyMix consistently outperformed competitive augmentation methods across various domain transfer settings. The dynamic adjustment of frequency regions enables DyMix to better handle domain shifts, especially leading to improved ACC and AUC scores. This ability highlighted DyMix's efficiency in enhancing the model's ability to generalize across different datasets and clinical settings.

\begin{table}[t!]\centering\scriptsize\setlength{\tabcolsep}{5pt}
    \caption{AD vs. CN Performance metrics ($\%$) of our proposed DyMix method compared with various data augmentation strategies during the domain adaptation step.}\label{tab_ab3}
    \begin{tabular}{cccccc}
        \toprule
        \text{Source} $\rightarrow$ \text{Target} & \textbf{Method} & \textbf{ACC} & \textbf{SEN} & \textbf{SPE} & \textbf{AUC}\\
        \midrule
        \multirow{6}{*}{\text{ADNI-1} $\rightarrow$ \text{ADNI-2}} & Mixup \cite{zhang2017mixup} & 88.89 & 84.37 & 92.50 & 88.44 \\
        & CutOut \cite{devries2017improved} & 86.11 & 71.87 & 97.50 & 84.69 \\
        & CutMix \cite{yun2019cutmix} & 90.27 & 84.37 & 95.00 & 89.68 \\
        & APR \cite{chen2021amplitude} & 90.27 & 81.25 & 97.50 & 89.37 \\
        & Fda \cite{yang2020fda} & 80.55 & \textbf{87.50} & 95.00 & 81.25 \\
        & \textbf{DyMix (Ours)} & \textbf{91.67} & 81.25 & \textbf{100.0} & \textbf{90.62} \\
        \midrule
        \multirow{6}{*}{\text{ADNI-1+ADNI-2} $\rightarrow$ \text{ADNI-3}} & Mixup \cite{zhang2017mixup} & 87.50 & 92.31 & 86.57 & 89.44 \\
        & CutOut \cite{devries2017improved} & 87.50 & 90.00 & 85.07 & 92.54 \\
        & CutMix \cite{yun2019cutmix} & 87.50 & 90.00 & 79.10 & 89.55 \\
        & APR \cite{chen2021amplitude} & 78.75 & 92.31 & 76.12 & 84.21 \\
        & Fda \cite{yang2020fda} & 88.75 & 92.31 & 88.06 & 90.18 \\
        & \textbf{DyMix (Ours)} & \textbf{91.25} & \textbf{92.31} & \textbf{89.10} & \textbf{95.70} \\
        \midrule
        \multirow{6}{*}{\text{ADNI-1} $\rightarrow$ \text{AIBL}} & Mixup \cite{zhang2017mixup} & 85.34 & 73.91 & 88.17 & 81.04 \\
        & CutOut \cite{devries2017improved} & 89.65 & 65.22 & 92.70 & 80.46 \\
        & CutMix \cite{yun2019cutmix} & 89.65 & 73.91 & 93.55 & 83.73 \\
        & APR \cite{chen2021amplitude} & 88.79 & 73.91 & 92.47 & 83.19 \\      
        & Fda \cite{yang2020fda} & 81.90 & 78.26 & 82.79 & 80.53 \\
        & \textbf{DyMix (Ours)} & \textbf{91.38} & \textbf{78.27} & \textbf{94.62} & \textbf{86.44} \\
        \midrule
        \multirow{6}{*}{\text{ADNI-1+ADNI-2} $\rightarrow$ \text{AIBL}} & Mixup \cite{zhang2017mixup} & 83.75 & 80.00 & 80.60 & 80.30\\
        & CutOut \cite{devries2017improved} & 79.31 & 78.26 & 79,57 & 78.92 \\
        & CutMix \cite{yun2019cutmix} & 86.21 & 69.56 & 90.32 & 79.94 \\
        & APR \cite{chen2021amplitude} & 87.93 & 69.56 & 90.47 & 80.02 \\  
        & Fda \cite{yang2020fda} & 80.00 & 80.00 & 76.12 & 80.55 \\
        & \textbf{DyMix(Ours)} & \textbf{89.31} & \textbf{82.61} & \textbf{90.65} & \textbf{88.06} \\
        \midrule
        \multirow{6}{*}{\text{AIBL} $\rightarrow$ \text{ADNI-3}} & Mixup \cite{zhang2017mixup} & 74.14 & 52.61 & 72.04 & 77.32\\
        & CutOut \cite{devries2017improved} & 86.25 & 53.85 & 92.54 & 73.19 \\
        & CutMix \cite{yun2019cutmix} & 86.25 & 61.54 & 91.04 & 76.29 \\
        & APR \cite{chen2021amplitude} & 83.75 & 61.54 & 88.06 & 74.80 \\  
        & Fda \cite{yang2020fda} & 80.17 & \textbf{78.26} & 80.64 & 79.45 \\
        & \textbf{DyMix (Ours)} & \textbf{91.25} & 53.85 & \textbf{98.51} & \textbf{80.18} \\
        \bottomrule
    \end{tabular}
\end{table}

\section{Conclusion}
\label{sec:conclusion}
In this study, we introduce a novel DyMix technique for the UDA approach in the context of AD diagnosis. We have shown that our proposed method addresses the challenges posed by domain shifts, which are common in medical imaging, by mitigating the non-uniform data distribution gap between the source and target domains. In contrast to conventional UDA methods that primarily focus on aligning local features or rely on fixed frequency manipulations, DyMix dynamically adjusts the mixing regions in the frequency domain, optimizing the model’s ability to adapt to domain variability and improving generalization across unseen data. Additionally, we enhanced the model’s resilience to intensity variations by combining amplitude-phase recombination and self-adversarial learning with spatial attention to produce invariant feature representations during the pretraining phase. In this way, the model not only adapts well to new domains but also maintains high diagnostic accuracy and reliability.

Rigorous evaluation regimens that included qualitative investigations and quantitative comparisons validated on two benchmark datasets (\ie, the ADNI and AIBL datasets) demonstrated that DyMix consistently outperformed state-of-the-art UDA methods across multiple domain transfer scenarios. Compared to other frequency-based approaches, we further verified that our method showed substantial improvements in all domain transfer scenarios, highlighting its effectiveness in handling domain shifts and enhancing AD diagnosis.

In summary, exploiting the DyMix technique offers a robust and adaptive solution for domain adaptation in medical imaging, particularly for AD diagnosis, where domain variability poses a significant challenge. In this light, the future direction of our work will focus on extending this framework to other neurodegenerative diseases and exploring its applicability to different imaging modalities, such as functional MRI and computed tomography. Additionally, we believe that integrating more advanced dynamic scheduling strategies and further refining the frequency-based mixup technique could provide additional improvements and broaden the method's impact in clinical applications.

\bibliographystyle{IEEEtran}
\bibliography{TMI/main}

\begin{thebibliography}{10}
\providecommand{\url}[1]{#1}
\csname url@samestyle\endcsname
\providecommand{\newblock}{\relax}
\providecommand{\bibinfo}[2]{#2}
\providecommand{\BIBentrySTDinterwordspacing}{\spaceskip=0pt\relax}
\providecommand{\BIBentryALTinterwordstretchfactor}{4}
\providecommand{\BIBentryALTinterwordspacing}{\spaceskip=\fontdimen2\font plus
\BIBentryALTinterwordstretchfactor\fontdimen3\font minus \fontdimen4\font\relax}
\providecommand{\BIBforeignlanguage}[2]{{%
\expandafter\ifx\csname l@#1\endcsname\relax
\typeout{** WARNING: IEEEtran.bst: No hyphenation pattern has been}%
\typeout{** loaded for the language `#1'. Using the pattern for}%
\typeout{** the default language instead.}%
\else
\language=\csname l@#1\endcsname
\fi
#2}}
\providecommand{\BIBdecl}{\relax}
\BIBdecl

\bibitem{frisoni2010clinical}
G.~B. Frisoni, N.~C. Fox, C.~R. Jack~Jr, P.~Scheltens, and P.~M. Thompson, ``The clinical use of structural mri in alzheimer disease,'' \emph{Nature Reviews Neurology}, vol.~6, no.~2, pp. 67--77, 2010.

\bibitem{brookmeyer2007forecasting}
R.~Brookmeyer, E.~Johnson, K.~Ziegler-Graham, and H.~M. Arrighi, ``Forecasting the global burden of alzheimer’s disease,'' \emph{Alzheimer's \& Dementia}, vol.~3, no.~3, pp. 186--191, 2007.

\bibitem{alzheimer20192019}
A.~Association, ``2019 alzheimer's disease facts and figures,'' \emph{Alzheimer's \& dementia}, vol.~15, no.~3, pp. 321--387, 2019.

\bibitem{zhao2023conventional}
Z.~Zhao, J.~H. Chuah, K.~W. Lai, C.-O. Chow, M.~Gochoo, S.~Dhanalakshmi, N.~Wang, W.~Bao, and X.~Wu, ``Conventional machine learning and deep learning in alzheimer's disease diagnosis using neuroimaging: A review,'' \emph{Frontiers in Computational Neuroscience}, vol.~17, p. 1038636, 2023.

\bibitem{khan2021machine}
P.~Khan, M.~F. Kader, S.~R. Islam, A.~B. Rahman, M.~S. Kamal, M.~U. Toha, and K.-S. Kwak, ``Machine learning and deep learning approaches for brain disease diagnosis: principles and recent advances,'' \emph{IEEE Access}, vol.~9, pp. 37\,622--37\,655, 2021.

\bibitem{zhang2020survey}
L.~Zhang, M.~Wang, M.~Liu, and D.~Zhang, ``A survey on deep learning for neuroimaging-based brain disorder analysis,'' \emph{Frontiers in Neuroscience}, vol.~14, p. 779, 2020.

\bibitem{ben2006analysis}
S.~Ben-David, J.~Blitzer, K.~Crammer, and F.~Pereira, ``Analysis of representations for domain adaptation,'' \emph{Advances in Neural Information Processing Systems}, vol.~19, 2006.

\bibitem{wilson2020survey}
G.~Wilson and D.~J. Cook, ``A survey of unsupervised deep domain adaptation,'' \emph{ACM Transactions on Intelligent Systems and Technology}, vol.~11, no.~5, pp. 1--46, 2020.

\bibitem{ganin2016domain}
Y.~Ganin, E.~Ustinova, H.~Ajakan, P.~Germain, H.~Larochelle, F.~Laviolette, M.~March, and V.~Lempitsky, ``Domain-adversarial training of neural networks,'' \emph{Journal of Machine Learning Research}, vol.~17, no.~59, pp. 1--35, 2016.

\bibitem{sun2017correlation}
B.~Sun, J.~Feng, and K.~Saenko, ``Correlation alignment for unsupervised domain adaptation,'' \emph{Domain Adaptation in Computer Vision Applications}, pp. 153--171, 2017.

\bibitem{guan2021multi}
H.~Guan, Y.~Liu, E.~Yang, P.-T. Yap, D.~Shen, and M.~Liu, ``Multi-site {MRI} harmonization via attention-guided deep domain adaptation for brain disorder identification,'' \emph{Medical Image Analysis}, vol.~71, p. 102076, 2021.

\bibitem{cai2023prototype}
H.~Cai, Q.~Zhang, and Y.~Long, ``Prototype-guided multi-scale domain adaptation for alzheimer's disease detection,'' \emph{Computers in Biology and Medicine}, vol. 154, p. 106570, 2023.

\bibitem{nussbaumer1982fast}
H.~J. Nussbaumer and H.~J. Nussbaumer, \emph{The fast Fourier transform}.\hskip 1em plus 0.5em minus 0.4em\relax Springer, 1982.

\bibitem{yang2020fda}
Y.~Yang and S.~Soatto, ``Fda: Fourier domain adaptation for semantic segmentation,'' in \emph{Proceedings of the IEEE/CVF Conference on Computer Vision and Pattern Recognition}, 2020, pp. 4085--4095.

\bibitem{hu2022domain}
S.~Hu, Z.~Liao, and Y.~Xia, ``Domain specific convolution and high frequency reconstruction based unsupervised domain adaptation for medical image segmentation,'' in \emph{International Conference on Medical Image Computing and Computer-Assisted Intervention}.\hskip 1em plus 0.5em minus 0.4em\relax Springer, 2022, pp. 650--659.

\bibitem{ge2023unsupervised}
Y.~Ge, Z.-M. Chen, G.~Zhang, A.~A. Heidari, H.~Chen, and S.~Teng, ``Unsupervised domain adaptation via style adaptation and boundary enhancement for medical semantic segmentation,'' \emph{Neurocomputing}, vol. 550, p. 126469, 2023.

\bibitem{oh2024fiesta}
K.~Oh, E.~Jeon, D.-W. Heo, Y.~Shin, and H.-I. Suk, ``Fiesta: Fourier-based semantic augmentation with uncertainty guidance for enhanced domain generalizability in medical image segmentation,'' \emph{arXiv preprint arXiv:2406.14308}, 2024.

\bibitem{shin2023frequency}
Y.~Shin, J.~Maeng, K.~Oh, and H.-I. Suk, ``Frequency mixup manipulation based unsupervised domain adaptation for brain disease identification,'' in \emph{Asian Conference on Pattern Recognition}.\hskip 1em plus 0.5em minus 0.4em\relax Springer, 2023, pp. 123--135.

\bibitem{zhang2017mixup}
H.~Zhang, M.~Cisse, Y.~N. Dauphin, and D.~Lopez-Paz, ``mixup: Beyond empirical risk minimization,'' \emph{arXiv preprint arXiv:1710.09412}, 2017.

\bibitem{chen2021amplitude}
G.~Chen, P.~Peng, L.~Ma, J.~Li, L.~Du, and Y.~Tian, ``Amplitude-phase recombination: Rethinking robustness of convolutional neural networks in frequency domain,'' in \emph{Proceedings of the IEEE/CVF International Conference on Computer Vision}, 2021, pp. 458--467.

\bibitem{zhou2023self}
Q.~Zhou, Q.~Gu, J.~Pang, X.~Lu, and L.~Ma, ``Self-adversarial disentangling for specific domain adaptation,'' \emph{IEEE Transactions on Pattern Analysis and Machine Intelligence}, 2023.

\bibitem{mueller2005alzheimer}
S.~G. Mueller, M.~W. Weiner, L.~J. Thal, R.~C. Petersen, C.~Jack, W.~Jagust, J.~Q. Trojanowski, A.~W. Toga, and L.~Beckett, ``The alzheimer’s disease neuroimaging initiative,'' \emph{Neuroimaging Clinics of North America}, vol.~15, no.~4, p. 869, 2005.

\bibitem{rowe2010amyloid}
C.~C. Rowe, K.~A. Ellis, M.~Rimajova, P.~Bourgeat, K.~E. Pike, G.~Jones, J.~Fripp, H.~Tochon-Danguy, L.~Morandeau, G.~O'Keefe \emph{et~al.}, ``Amyloid imaging results from the australian imaging, biomarkers and lifestyle (aibl) study of aging,'' \emph{Neurobiology of Aging}, vol.~31, no.~8, pp. 1275--1283, 2010.

\bibitem{perez2021torchio}
F.~P{\'e}rez-Garc{\'\i}a, R.~Sparks, and S.~Ourselin, ``Torchio: a python library for efficient loading, preprocessing, augmentation and patch-based sampling of medical images in deep learning,'' \emph{Computer Methods and Programs in Biomedicine}, vol. 208, p. 106236, 2021.

\bibitem{lian2018hierarchical}
C.~Lian, M.~Liu, J.~Zhang, and D.~Shen, ``Hierarchical fully convolutional network for joint atrophy localization and alzheimer's disease diagnosis using structural mri,'' \emph{IEEE transactions on pattern analysis and machine intelligence}, vol.~42, no.~4, pp. 880--893, 2018.

\bibitem{mu2011adult}
Y.~Mu and F.~H. Gage, ``Adult hippocampal neurogenesis and its role in alzheimer's disease,'' \emph{Molecular neurodegeneration}, vol.~6, pp. 1--9, 2011.

\bibitem{woo2018cbam}
S.~Woo, J.~Park, J.-Y. Lee, and I.~S. Kweon, ``Cbam: Convolutional block attention module,'' in \emph{Proceedings of the European conference on computer vision (ECCV)}, 2018, pp. 3--19.

\bibitem{isensee2019automated}
F.~Isensee, M.~Schell, I.~Pflueger, G.~Brugnara, D.~Bonekamp, U.~Neuberger, A.~Wick, H.-P. Schlemmer, S.~Heiland, W.~Wick \emph{et~al.}, ``Automated brain extraction of multisequence mri using artificial neural networks,'' \emph{Human Brain Mapping}, vol.~40, no.~17, pp. 4952--4964, 2019.

\bibitem{zhang2001segmentation}
Y.~Zhang, M.~Brady, and S.~Smith, ``Segmentation of brain mr images through a hidden markov random field model and the expectation-maximization algorithm,'' \emph{IEEE Transactions on Medical Imaging}, vol.~20, no.~1, pp. 45--57, 2001.

\bibitem{kingma2014adam}
D.~P. Kingma and J.~Ba, ``Adam: A method for stochastic optimization,'' \emph{arXiv preprint arXiv:1412.6980}, 2014.

\bibitem{selvaraju2020grad}
R.~R. Selvaraju, M.~Cogswell, A.~Das, R.~Vedantam, D.~Parikh, and D.~Batra, ``Grad-cam: visual explanations from deep networks via gradient-based localization,'' \emph{International Journal of Computer Vision}, vol. 128, pp. 336--359, 2020.

\bibitem{risacher2009baseline}
S.~L. Risacher, A.~J. Saykin, J.~D. Wes, L.~Shen, H.~A. Firpi, and B.~C. McDonald, ``Baseline mri predictors of conversion from mci to probable ad in the adni cohort,'' \emph{Current Alzheimer Research}, vol.~6, no.~4, pp. 347--361, 2009.

\bibitem{davies1987quantitative}
C.~Davies, D.~Mann, P.~Sumpter, and P.~Yates, ``A quantitative morphometric analysis of the neuronal and synaptic content of the frontal and temporal cortex in patients with alzheimer's disease,'' \emph{Journal of the neurological sciences}, vol.~78, no.~2, pp. 151--164, 1987.

\bibitem{devries2017improved}
T.~DeVries and G.~W. Taylor, ``Improved regularization of convolutional neural networks with cutout,'' \emph{arXiv preprint arXiv:1708.04552}, 2017.

\bibitem{yun2019cutmix}
S.~Yun, D.~Han, S.~J. Oh, S.~Chun, J.~Choe, and Y.~Yoo, ``Cutmix: Regularization strategy to train strong classifiers with localizable features,'' in \emph{Proceedings of the IEEE/CVF international conference on computer vision}, 2019, pp. 6023--6032.

\end{thebibliography}

\end{document}